\documentclass[a4paper,11pt]{article}
\pdfoutput=1
\usepackage{graphicx}
\usepackage{amsmath}
\usepackage{mathrsfs}
\usepackage{array}
\graphicspath{{./figures/}}
\usepackage{jinstpub}
\AtBeginDocument{\mathcode`v=\varv}

\title{On determining the fraction of metastable ions produced by direct ionization}

\author[1]{F. Chu,\note{Corresponding author.}}

\author{S.W. Mattingly, J. Berumen, R. Hood and F. Skiff}

\affiliation{Department of Physics and Astronomy, University of Iowa,\\Iowa City, IA 52242, USA}

\emailAdd{feng-chu@uiowa.edu}

\abstract{Laser-induced fluorescence (LIF) is a powerful tool in studying wave-particle interactions, velocity-space diffusion and other phenomena in plasmas under the proper conditions. Ignoring the possible instrumental errors in LIF, such as metastable lifetime effects, may result in unreliable measurements. LIF is frequently performed on metastable states that are produced from direct ionization of neutral gas particles and ions in other electronic states. However, the metastable population born from neutrals cannot faithfully represent processes which act on the ion dynamics in a time shorter than the metastable lifetime. A numerical simulation is performed to study the metastable lifetime effects using a Lagrangian approach for LIF. A theoretical model in determining the fraction of metastable ions produced from direct ionization is reported to provide corrections to the LIF measurements.}


\proceeding{18$^{\text{th}}$ International Symposium on Laser-Aided Plasma Diagnostics,\\24--28 September 2017,\\Prague, Czech Republic}

\begin{document}
\maketitle
\flushbottom

\section{Introduction}
\label{sec:intro}

Laser-induced fluorescence (LIF) is a nonintrusive, nominally nonperturbative plasma diagnostic technique. In the case where the effect of photon momentum on ion orbits is negligible, LIF provides an important window into the dynamics of ion motions. A reliable phase-space diagnostic is required in the study of plasma electrostatic degrees of freedom \cite{van_kampen_theory_1955, case_plasma_1959, skiff_electrostatic_2002}, ion heating \cite{anderegg_ion_1986}, velocity-space diffusion \cite{curry_measurement_1995} and related phenomena \cite{skiff_direct_1987} in gas discharge, fusion, and other plasmas.

For practical purposes, LIF is frequently performed on metastable states that are produced from neutral gas particles \cite{cherrington_gaseous_1979} and ions in other electronic states. LIF is observed from allowed transitions of plasma ions that are optically pumped to excited states. This optical pumping process, which plays a key role in LIF, is dependent on ion orbits and ultimately is non-linear in laser intensity. If one tries to increase the laser intensity to obtain a better signal, then optical pumping can produce systematic errors in LIF measurements known as optical pumping broadening \cite{goeckner_laserinduced_1989, goeckner_saturation_1993}. 

Another type of instrumental effects that have not been well studied are caused by the lifetime of metastable ions. One important consequence of the metastable lifetime effects is that under circumstances where the metastable ion population is produced from direct ionization of neutrals (as opposed to the excitation of ground-state ions), the ion velocity distribution $f_0(v)$ and its perturbation $f_1(v,t)$ measured using LIF will only faithfully represent processes which act on the ion dynamics in a time shorter than the metastable lifetime \cite{cooper}. For example, the ion temperature measurements are only accurate when the metastable ions can live longer than the ion-ion collision mean free time. Similarly, in the wave measurements, the wave period has to be significantly shorter than the metastable lifetime for a direct interpretation. However, the contribution to the LIF signal from the metastable population produced from pre-existing ions is much less complicated. Since these metastables have a history as ``typical'' ions and have already represented the actual ion distribution, the measured full widths of the velocity distribution function is independent of metastable lifetime. Similarly, given that these metastable ions have started to respond to the wave field long time before they are produced, the measured $f_1(v,t)$ is also not affected by the metastable lifetime effects.

Even though the electronic cross section for direct electron-impact production of metastable ions from neutrals is significantly smaller than the excitation cross section that produces metastables from ground-state ions \cite{goeckner_laserinduced_1991}, the density of neutrals can be orders of magnitude higher than the ground-state ion density. As a result, considerable metastable ions can be produced from direct ionization of neutrals and introduce the systematic errors due to metastable lifetime effects. Therefore, an accurate model for determining the fraction of this metastable population is needed to provide corrections to the LIF measurements. This paper is organized as follows: section \ref{sec:LIF} presents a typical LIF scheme and rate equations, section \ref{sec:theo} gives a description of the Lagrangian approach for LIF, section \ref{sec:results} presents the simulation result and the theoretical model in determining the fraction of metastables produced from direct ionization, and section \ref{sec:summary} provides a summary.

\section{LIF Scheme and Rate Equations}
\label{sec:LIF}

\begin{figure}
\begin{center}
\includegraphics[width=3.1in]{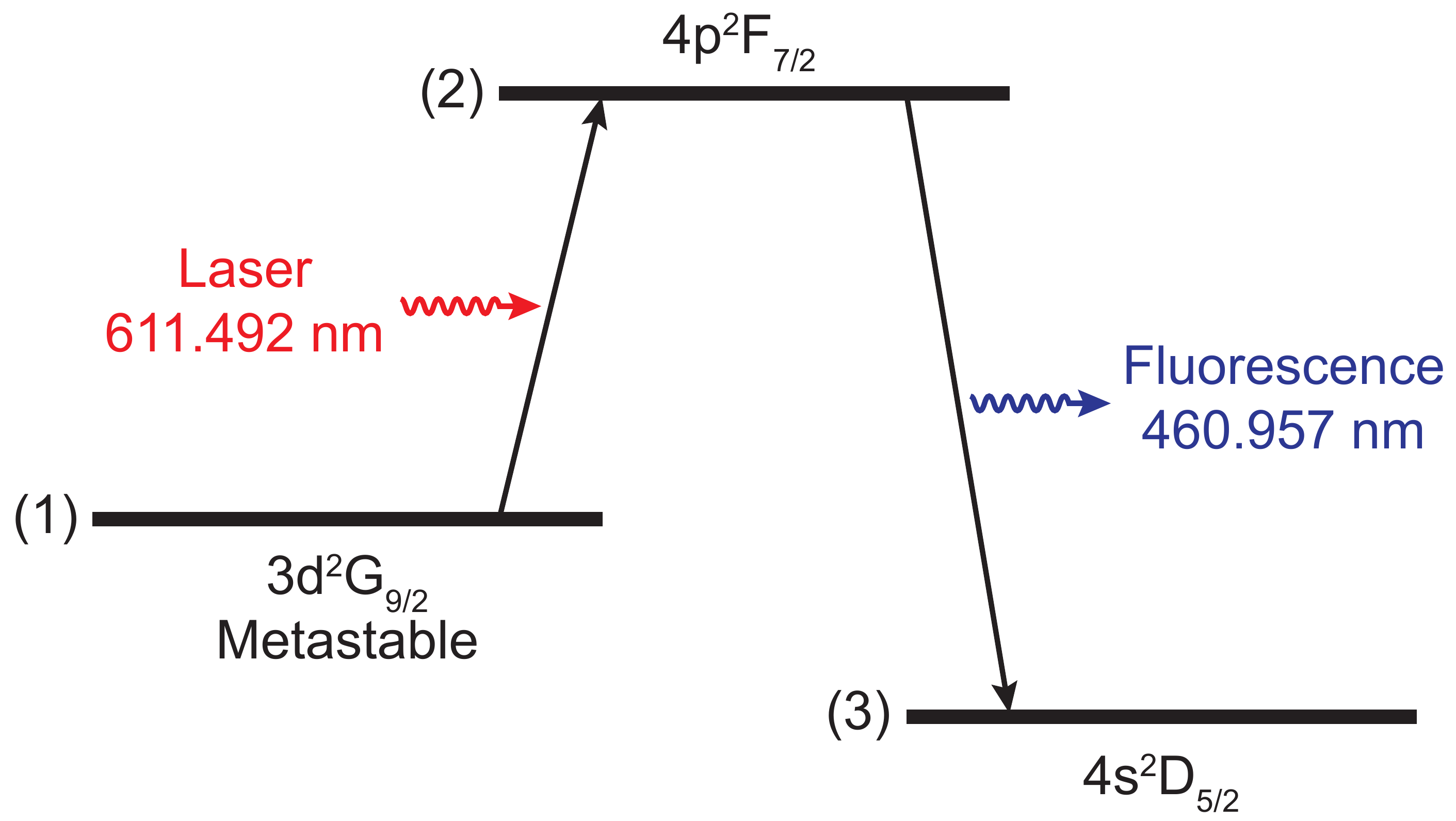}
\caption{Typical energy level diagram for LIF in ArII. To induce fluorescence, in the rest frame of an ion, a laser is tuned at 611.492 nm to excite electrons in metastable state 1 to excited state 2. The electrons in state 2 then spontaneously decay to state 3, emitting fluorescence photons at 460.957 nm.}
\label{fig:LIF}
\end{center}
\end{figure}

Laser-induced fluorescence is typically described using a three-level system. A commonly used energy level diagram for ArII \cite{mattingly_measurement_2013} is shown in figure~\ref{fig:LIF}. One starts with an ion in the $\textup{3d}^2\textup{G}_{9/2}$ metastable state, and a laser is tuned at 611.492 nm to excite electrons to the excited state $\textup{4p}^2\textup{F}_{7/2}$. Fluorescence photons are emitted at 460.957 nm when those electrons decay to the $\textup{4s}^2\textup{D}_{5/2}$ state with a branching ratio of $66.5\%$ \cite{severn_argon_1998}. The set of rate equations for states 1, 2, and 3 in figure \ref{fig:LIF} are:
\begin{subequations}
\begin{align}
\label{eq:rate1}
\frac{dn_1}{dt}&=-(w+r+u)n_1+A_{21}n_2,\\
\label{eq:rate2}
\frac{dn_2}{dt}&=-A_\textup{T}n_2+(w+u)n_1,\\
\label{eq:rate3}
\frac{dn_3}{dt}&=A_{23}n_2,
\end{align}
\end{subequations}
where $r$ is the metastable quench rate, $u$ is the electron-collisional excitation rate, $A_{ij}$ is the Einstein coefficient of spontaneous emission, $A_\textup{T}$ is the total spontaneous decay rate of the excited state, and $w$ is the optical pumping rate. The probability that an ion is in level 1 and 2 is denoted by $n_1$ and $n_2$ respectively, while $n_3$ is the probability that the ion spontaneously decays from level 2 to level 3. Apparently, the initial conditions are $n_1=1$ and $n_2=n_3=0$. Stimulated emission can be ignored here because it is smaller than the spontaneous decay rates even in the relatively strong optical pumping regime.

\begin{figure}
\begin{center}
\includegraphics[width=3.2in]{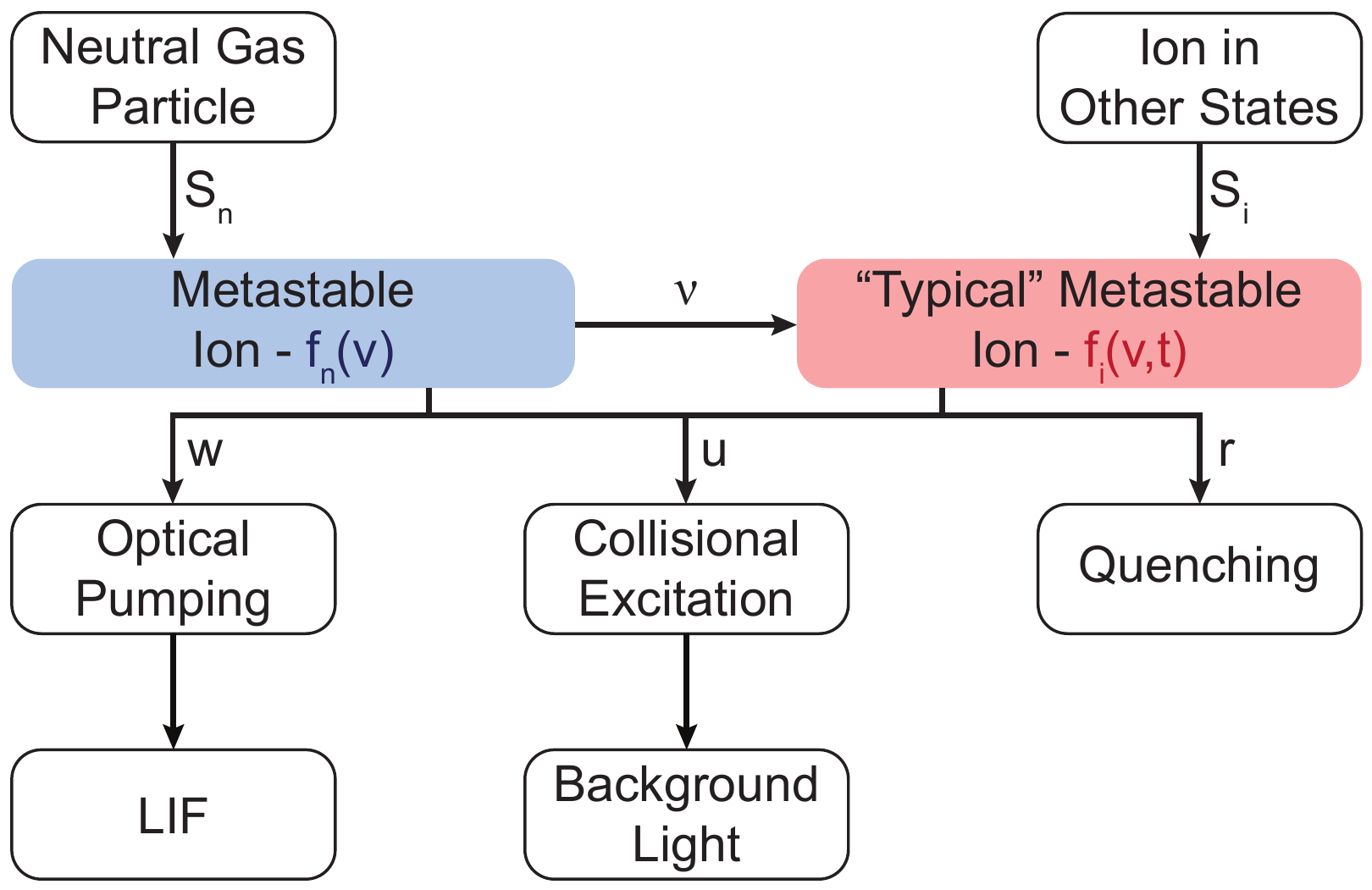}
\caption{Schematic illustration of the production and loss mechanisms of a metastable ion. $S_\textup{n}$ is the metastable birth rate from direct ionization of a neutral gas particle and $S_\textup{i}$ is birth rate from an ion in other electronic states. $f_\textup{n}(v)$ and $f_\textup{i}(v,t)$ are the neutral and ion velocity distribution functions respectively. $\nu$ is the coulomb collision frequency between ions.} 
\label{fig:dia}
\end{center}
\end{figure}

The production and loss mechanisms of a metastable ion are shown schematically in figure~\ref{fig:dia}. Metastables can be produced from both neutral gas particles and pre-existing ions. Metastable born from the latter has a history as a ``typical'' ion. However, the metastable coming from direct ionization is initially representative of the neutral velocity distribution and only becomes ``typical'' over time through ion-ion coulomb collisions. Once a metastable is produced, it can be lost primarily through three mechanisms: optical pumping, collisional excitation and quenching. The first two contribute to the LIF signal and background fluorescence light respectively.

\section{1-D Lagrangian Model for LIF}
\label{sec:theo}

Laser-induced fluorescence is normally analyzed through an Eulerian approach, where the rate equations are extended to a system of coupled kinetic equations, with one for each quantum state. The solutions of the coupled kinetic equations, being a set of coupled partial differential equations (PDEs), become extremely difficult to compute with the existence of nonuniformity in phase space, e.g., an electrostatic wave. Alternatively, a Lagrangian approach for LIF is adopted to achieve large computational advantages by exploiting the separation of the classical dynamics of the ions from the quantum mechanics of the electronic states.

A Lagrangian approach starts with a description of the ion orbits in phase space. A conditional probability function $P (\mathbf{x},t;\mathbf{x'},t')$ is introduced to specify the probability of finding an ion at the phase-space point $\mathbf{x}=(x,v)$ at time $t$, given that the ion was at point $\mathbf{x'}=(x',v')$ at time $t'$. This function is the Green's function of the kinetic equation for the ions \cite{dougherty_model_1964, chandrasekhar_stochastic_1943}. Adopting a simple 1-D Fokker-Planck model with a constant ion-ion coulomb collision frequency $\nu$, the function $P$ is given by
\begin{equation}
\label{eq:P}
P (\mathbf{x},t;\mathbf{x'},t')=p(t-t')\exp \left[-\frac{1}{2}\left ( \mathbf{x}-\mathbf{\tilde{x}} \right ) \mathbf{q} \left(\mathbf{x}-\mathbf{\tilde{x}} \right)^{\mathsf{T}} \right]
\end{equation}
where $\mathbf{\tilde{x}}(t)=(\tilde{x},\tilde{v})$ is the ion orbit in the absence of velocity-space diffusion starting at $\mathbf{\tilde{x}}(0)=\mathbf{x'}$. To simplify the calculations to the first order, the electric field $\mathbf{E}$ and magnetic field $\mathbf{B}$ are included in the ion orbit $\mathbf{\tilde{x}}(t)$ instead of the kinetic equation. The matrix $\mathbf{q}$ is defined as given by
\begin{equation}
\label{eq:q}
\mathbf{q}=\begin{pmatrix}
\frac{\nu ^3\left (1+e^{\nu t} \right )}{2\eta \left [ e^{\nu t}(\nu t-2)+\nu t+2 \right ]} &
\frac{\nu ^2\left (1-e^{\nu t} \right )}{2\eta \left [ e^{\nu t}(\nu t-2)+\nu t+2 \right ]} \\[14pt]
\frac{\nu ^2\left (1-e^{\nu t} \right )}{2\eta \left [ e^{\nu t}(\nu t-2)+\nu t+2 \right ]} & 
\frac{\nu \left [ e^{2\nu t}(2\nu t-3)+4e^{\nu t}-1 \right ]}{2\eta \left(e^{\nu t}-1 \right) \left [ e^{\nu t}(\nu t-2)+\nu t+2 \right ]} 
\end{pmatrix},
\end{equation}
and $p(t)$ is given by
\begin{equation}
\label{eq:p(t)}
p(t)=\frac{\nu ^2 e^{\nu t}}{2 \sqrt{2} \pi \eta }\left \{\left ( e^{\nu t}-1 \right )\left [ \nu t+e^{\nu t}(\nu t-2)+2 \right ] \right \}^{-\frac{1}{2}},
\end{equation}
where $\eta=\nu T_\textup{i}/m_\textup{i}$, $T_\textup{i}$ is the ion temperature in energy units, and $m_\textup{i}$ is the ion mass.

The next step in the Lagrangian approach is to compute optical pumping, which is a function on ion orbits. For metastable ions with velocity $v$ and single-frequency laser intensity $I$, the optical pumping rate for a given transition can be expressed as
\begin{equation}
\label{eq:w}
w(v,t)=\frac{BI(t)}{c\gamma }g(v-v_\textup{L}),
\end{equation}
where $B$ is the Einstein coefficient of induced absorption, $\gamma$ is the half linewidth of the transition in frequency units, $g$ is a dimensionless Lorentzian function that represents the natural line shape of the transition in velocity units, and $v_\textup{L}$ is the velocity that an ion must have to Doppler shift the laser light into resonance with the transition. The line shape $g$ is
\begin{equation}
\label{eq:g}
g(v-v_\textup{L})=\frac{1}{\pi}\cdot \frac{\left ( \gamma \lambda _\textup{L} \right )^{2}}{\left ( v-v_\textup{L} \right )^2+\left ( \gamma \lambda _\textup{L} \right )^2} ,
\end{equation}
where $\lambda_\textup{L}$ is the laser wavelength. Taking the velocity-space diffusion into account, the total optical pumping rate for a metastable ion at time $t$ is calculated by averaging over the conditional probability function $P$:
\begin{equation}
\label{eq:Wn}
W(t;\mathbf{x'},t' )=\int w\left (v,t \right )P\left ( \mathbf{x},t;\mathbf{x'},t' \right )d\mathbf{x}.
\end{equation}
Equation (\ref{eq:Wn}) is a convolution of a Lorentzian and a Gaussian profile, i.e., a Voigt function, which can be evaluated analytically through a continued fraction expansion of the plasma dispersion function \cite{armstrong_spectrum_1967,mccabe_continued_1984}.

With the help of the average optical pumping rate $W$ and equation (\ref{eq:rate1}), the quantum state probability $\Psi$ that an ion remains in metastable state at time $t$ can be calculated as
\begin{equation}
\label{eq:psi}
\Psi (t;\mathbf{x'},t')=\exp \left\{ - \left [ \left (1-\frac{A_{21}}{A_\textup{T}} \right ) \int_{t'}^{t}W \left ( t''; \mathbf{x'},t' \right )dt'' +\left (1-\frac{A_{21}}{A_\textup{T}} \right )u\left ( t-t' \right )+r\left ( t-t' \right ) \right ] \right \},
\end{equation}
where the initial value $\Psi (t';\mathbf{x'},t')$ is unity. Therefore, the total probability distribution of finding a metastable ion at its final position $\mathbf{x}$ in phase-space at time $t$ after it was produced at the initial position $\mathbf{x'}$ at time $t'$ can be constructed by combining equations (\ref{eq:P}) and (\ref{eq:psi}):
\begin{equation}
\label{eq:Q}
Q (\mathbf{x},t;\mathbf{x'},t')=P (\mathbf{x},t;\mathbf{x'},t') \Psi (t;\mathbf{x'},t').
\end{equation}

In the three-level system described in figure  \ref{fig:LIF}, every observable fluorescence photon comes from an optically pumped ion that decays from level 2 to level 3. According to equation (\ref{eq:rate3}), the contribution of a metastable ion produced at the initial phase-space point $\mathbf{x'}$ at time $t'$ to the fluorescence signal (in photons/s) in the LIF viewing volume is
\begin{equation}
\label{eq:signal}
N (t;\mathbf{x'},t')=\frac{A_{23}}{A_\textup{T}}\iint\left [w(v,t)+u \right ]Q(\mathbf{x},t;\mathbf{x'},t')\theta (x)d\mathbf{x},
\end{equation}
where $\theta (x)$ is a normalized window function.

The last step in the Lagrangian approach is to calculate the LIF signal produced from each of the metastable populations separately and then simply add them together. The contribution of the metastables born from direct ionization of neutrals is given by summing over all the initial conditions:
\begin{equation}
\label{eq:finalsignal1}
N_\textup{n} (t)=S_\textup{n}\iint\limits_{\mathbf{x'}}\int_{-\infty}^{t}N ( t;\mathbf{x'},t')f_\textup{n}(v')d\mathbf{x'}dt'.
\end{equation}
Similarly, the contribution of the metastables coming from pre-existing ions:
\begin{equation}
\label{eq:finalsignal2}
N_\textup{i} (t)=S_\textup{i}\iint\limits_{\mathbf{x'}}\int_{-\infty}^{t}N ( t;\mathbf{x'},t')f_\textup{i}(v',t')d\mathbf{x'}dt'.
\end{equation}
In equations (\ref{eq:finalsignal1}) and (\ref{eq:finalsignal2}), $S_\textup{n}f_\textup{n}(v)$ is the metastable birth rate from neutrals and $S_\textup{i}f_\textup{i}(v,t)$ is the metastable birth rate from pre-existing ions at time $t$. Therefore, the total LIF signal is given by
\begin{equation}
\label{eq:finalsignal}
N (t)=N_\textup{n} (t)+N_\textup{i} (t).
\end{equation}

\section{Simulation Results}
\label{sec:results}

Based on the Lagrangian approach for LIF above, a numerical simulation is performed to study how the perturbed distribution function $f_1$ measured using LIF changes with the metastable lifetime. Function $f_1$ is obtained by scanning the laser wavelength and analyzing the Fourier transform of the LIF signal in equation (\ref{eq:finalsignal}) at different laser wavelengths. To simplify the simulation, we assume the plasma is only perturbed in the velocity-space. The ion orbit $\mathbf{\tilde{x}}(t)$ then becomes
\begin{equation}
\label{eq:orbit}
\dot{v}=-\nu v+\frac{Ee}{m_\textup{i}}\sin\left ( \omega t \right ),
\end{equation}
where the first term on the right side represents drag on the ion, and $E$ is the wave amplitude. As in many laboratory plasmas, the neutrals and ions are usually at the room temperature. Therefore, we also assume that the temperature of the neutrals is the same as ions in the simulation.

\subsection{Metastable lifetime effects}
\label{subsec:lifetime}

\begin{figure}
\begin{center}
\includegraphics[width=3.7in]{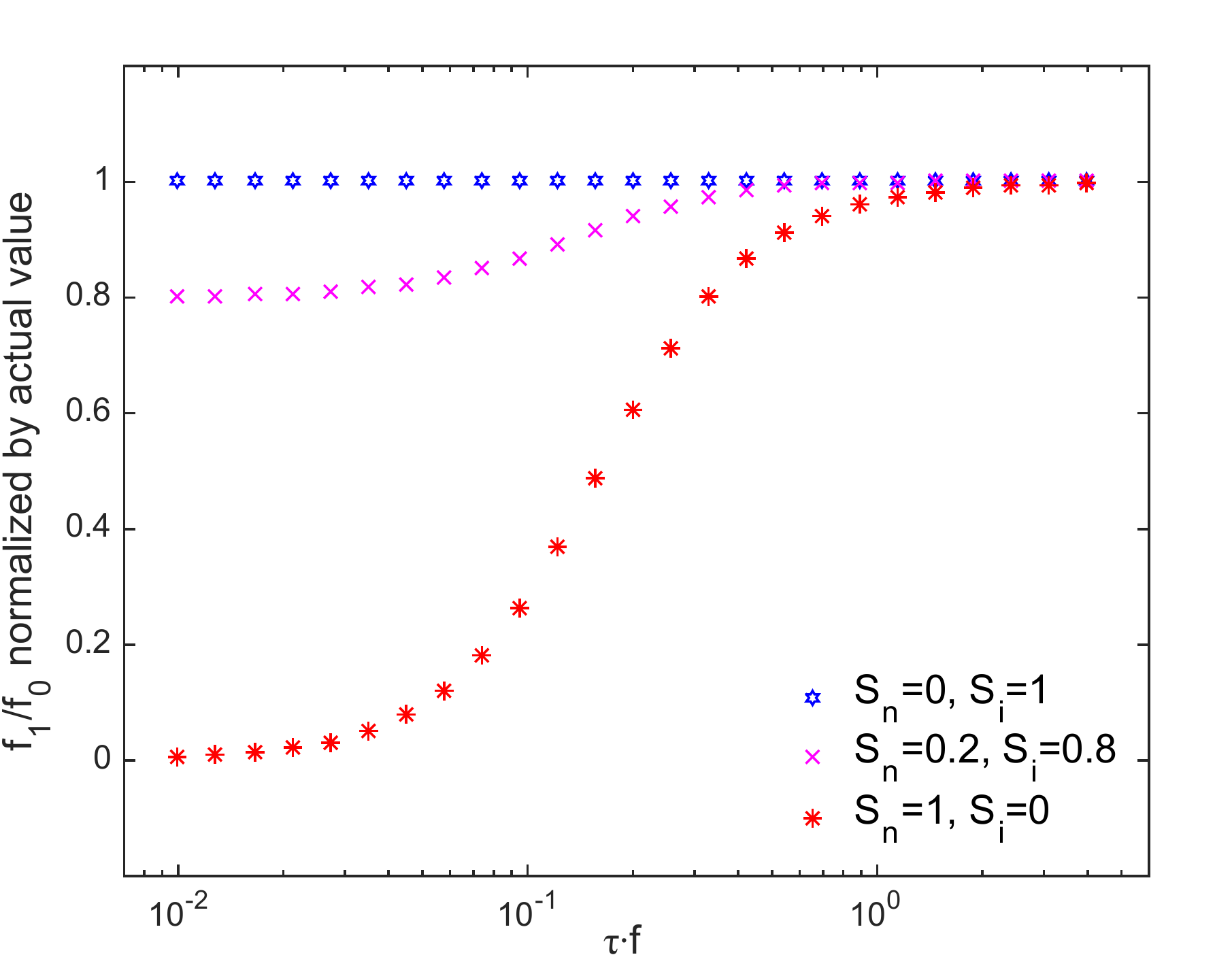}
\caption{Simulated ratio of $f_1$ to $f_0$ at $v=v_\textup{t}$ as a function of the metastable lifetime $\tau$ for different combinations of $S_\textup{n}$ and $S_\textup{i}$. Both $S_\textup{n}$ and $S_\textup{i}$ are normalized by the total birth rate $S=S_\textup{n}+S_\textup{i}$. $f$ is the frequency of the electrostatic wave. (1) $S_\textup{n}=0$ and $S_\textup{i}=1$. All the metastables are produced from pre-existing ions. $f_1/f_0$ always represents the correct ratio, no matter how long the metastable lifetime is. (2) $S_\textup{n}=0.2$ and $S_\textup{i}=0.8$. Metastables are produced from both neutral gas particles and pre-existing ions. The ratio $f_1/f_0$ is only valid when the metastable lifetime is longer than a wave period. This ratio is also bounded by two asymptotic limits when the metastable lifetime is much longer or shorter than a wave period. The ratio of the lower limit to the upper limit is 0.8. (3) $S_\textup{n}=1$ and $S_\textup{i}=0$. All the metastables are produced from direct ionization of neutrals. When the metastable lifetime is longer than a wave period, $f_1/f_0$ is valid. However, this ratio drops rapidly when the lifetime is shorter than a wave period.}
\label{fig:f1f0}
\end{center}
\end{figure}

When an electrostatic wave propagates in a plasma, the lifetime of metastable ions characterizes how long they typically experience the wave field. Figure \ref{fig:f1f0} shows the simulated ratio of $f_1$ to $f_0$ at $v=v_\textup{t}$ (thermal velocity) as a function of the metastable lifetime for different combinations of birth rates $S_\textup{n}$ and $S_\textup{i}$. For the metastables born from direct ionization of neutrals, they can only start to respond to the electric field once they are ionized. If the lifetime is too short compared to one wave period, these metastables will not live long enough to react to the wave, resulting in a reduction of the measured wave amplitude. On the other hand, the metastables produced from pre-existing ions are free of the metastable lifetime effects.

\subsection{Discussion}
\label{subsec:discussion}

As mentioned earlier, the metastable ions can be produced from both direct ionization of neutral gas particles and pre-existing ions. An interesting result about figure \ref{fig:f1f0} is that the ratio $f_1/f_0$ reaches an equilibrium when the metastable lifetime is either much longer or shorter than a wave period. In the former case, both of the metastable populations contribute to the LIF signal. However, in the latter situation, only the metastables born from pre-existing ions can effectively react to the electrostatic wave during their lifetime and contribute to the LIF signal. This fact provides a perfect opportunity to distinguish these two metastable populations by monitoring the LIF signal while scanning the frequency of the electrostatic wave and keeping its amplitude $E$ constant. The fraction of the metastables produced from direct ionization of neutrals can be determined by taking the ratio of the lower limit of the LIF signal to the upper limit. In practice, however, maintaining $E$ unchanged is a difficult task, since the antenna efficiency in launching the wave may vary with the frequency. An alternate approach is to measure the electrostatic wave amplitude using a Langmuir probe that is biased negatively to collect the ion saturation current. The LIF signal can then be normalized by the wave amplitude to avoid possible interference from the change in antenna efficiency.

\section{Summary}
\label{sec:summary}

In this paper, we report a theoretical model in determining the fraction of metastable ions produced from direct ionization. A numerical simulation based on the Lagrangian approach for LIF is performed to show that at higher frequencies LIF signal is produced from both of the metastable populations. However, at lower frequencies, only the metastables coming from pre-existing ions can contribute to the fluorescence signal. Through launching an electrostatic wave externally and scanning its frequency, the fraction of metastables produced from neutrals can be determined by taking the ratio of the lower limit of the LIF signal to the upper limit.

Experiments will be performed to study the metastable lifetime effects and test the theoretical model proposed above on one of the plasma chambers we have in Skiff's lab \cite{hood_ion_2016} at the University of Iowa.

\acknowledgments

This work was supported by the U.S. Department of Energy under Grant No. DE-FG02-99ER54543. This research is part of a Ph.D. dissertation to be submitted by F. Chu to the Graduate College, University of Iowa, Iowa City, IA.

\bibliography{refs}

\end{document}